\documentclass[twocolumn,showpacs,pre,floatfix,amsmath,amssymb,notitlepage]{revtex4-1}
\usepackage{epsfig}
\usepackage{subfigure}
\usepackage{amsmath}
\usepackage{color}
\usepackage{amssymb}
\usepackage{setspace}
\usepackage{graphicx}
\usepackage{dcolumn}
\usepackage{bm}
\usepackage{times}
\usepackage{enumerate}
\usepackage{footnote}
\input epsf

\newcommand{\appropto}{\mathrel{\vcenter{
  \offinterlineskip\halign{\hfil$##$\cr
    \propto\cr\noalign{\kern2pt}\sim\cr\noalign{\kern-2pt}}}}}

\graphicspath{{figures/}}

\begin{document}

\author{Per Sebastian Skardal}
\email{persebastian.skardal@trincoll.edu} 
\affiliation{Department of Mathematics, Trinity College, Hartford, CT 06106, USA}

\title{Quasi-periodic dynamics and a Neimark-Sacker bifurcation in nonlinear random walks on complex networks}

\begin{abstract}
We study the dynamics of nonlinear random walks on complex networks. We investigate the role and effect of directed network topologies on long-term dynamics. While a period-doubling bifurcation to alternating patterns occurs at a critical bias parameter value, we find that some directed structures give rise to a different kind of bifurcation that gives rise to quasi-periodic dynamics. This does not occur for all directed network structure, but only when the network structure is sufficiently directed. We find that the onset of quasi-periodic dynamics is the result of a Neimark-Sacker bifurcation, where a pair of complex-conjugate eigenvalues of the system Jacobian passes through the unit circle, destabilizing the stationary distribution with high-dimensional rotations. We investigate the nature of these bifurcations, study the onset of quasi-periodic dynamics as network structure is tuned to be more directed, and present an analytically tractable case of a four-neighbor ring.
\end{abstract}

\pacs{02.50.Ga,05.40.Fb,05.45.-a,89.75.Hc}

\maketitle

\section{Introduction}\label{sec:01}

Random walk dynamics have long been utilized for studying complex networks due to the vast information they provide with remarkably simple dynamics~\cite{Noh2002PRL,Masuda2017PR}. Examples where random walks and diffusion processes have proven useful in studying complex networks include Google's PageRank algorithm~\cite{Brin1998,Page1999,Gleich2015SIAM}, search and exploration~\cite{GomezGardenes2008PRE,Sinatra2011PRE}, modeling transport processes~\cite{Gorenflo2002,Nicosia2017PRL}, detection of communities and other network structures~\cite{Rosvall2008PNAS,Asllani2018PRL}, and finding geometric and topological embeddings~\cite{Coifman2005PNAS}. The simplicity of typical random walk dynamics follow from the Markovian, i.e., memory-less, nature of transitions and the fact that transition probabilities (or rates) are static. The result is a linear dynamical system where, assuming the relatively mild condition of a network structure being primitive, the dynamics converge to a unique, globally-attracting fixed point or stationary distribution~\cite{Durrett,MacCluer2000SIAM}. However, for the purpose of modeling more realistic scenarios where transition rules may not remain constant, the typical random walk model is too restrictive. For such cases we must relax the constraint on transition probabilities (or rates) being static, for which we expect to observe more complicated dynamics.

Here we study the dynamics of nonlinear random walks on complex networks~\cite{Skardal2019JNS}. Nonlinear random walks on network represent a recently formulated example of a nonlinear Markov process~\cite{Kolokoltsov2010,Frank2013} where classical dynamical systems techniques can be applied to study some remarkable dynamical behaviors. In particular, we consider a continuum of discrete-time random walkers on a network and let transition probabilities depend on the current state of the system. Using a bias parameter, random walkers may preferentially be biased towards neighboring nodes that are populated by relatively many or few other random walkers. This simple paradigm may be used to model the transport of resources in different scenarios, whereby individuals or institutions may prefer to allocate their resources to other wealthy or poor individuals or institutions under more capitalistic or humanitarian values, respectively. Such inequities in transition rules might also describe qualitative tendencies in migration patterns (i.e., individuals and families moving towards or away from more populated cities), employment patterns (i.e., individuals seeking to work at larger or smaller companies or institutions), and other scenarios where individuals or units are transported with various biases.

In previous work~\cite{Skardal2019JNS} it was shown that nonlinear random walks on networks consist of a weakly nonlinear regime where the bias is sufficiently small in magnitude as well as two strongly nonlinear regimes where the bias is more extreme and either positive or negative. In the weakly nonlinear regime it was proven that, as in the typical linear random, a unique, globally attracting fixed point or stationary distribution exists provided that the network structure is primitive. When the bias is sufficiently positive the steady-state dynamics localize strongly to a small subset of the networks' nodes and a number of such stable states emerge, with multistability becoming more pronounced as the bias becomes larger. Finally, for sufficiently negative bias a period-doubling bifurcation occurs when the network structure is undirected, beyond which the random walk converges to a period-two with relatively large oscillation amplitudes.

In this paper we focus on the strongly nonlinear regime with negative bias. In particular, we investigate the behavior of nonlinear random walks on networks that are more and more directed. As noted above, when the network structure is undirected a period-doubling bifurcation occurs where the stationary distribution loses stability and the random walk converges to a period-two orbit. However, when the underlying network structure is directed we observe more complicated dynamics in many cases. In such cases the system still exhibits a bifurcation where the stationary distribution loses stability, however beyond this point the dynamics no longer converge to a period-two orbit, but rather to a quasi-periodic orbit. We find that these dynamics stem from the nature of the bifurcation itself, which is no longer period-doubling, but rather a Neimark-Sacker bifurcation~\cite{Sacker2009,Kuznetsov2010}, marked by a pair of complex eigenvalues crossing the unit circle. Such quasi-periodic dynamics do not occur for all directed networks, but typically, only when the structure is ``sufficiently'' directed. That is, in typical cases a sufficiently large perturbation needs to be made to an existing undirected network structure to observe quasi-periodic dynamics, whether it be via rewiring or up/down-weighting enough links. However, in some cases the dynamics become quasi-periodic with arbitrarily small ``directedness'', as we see in an example with a four-neighbor ring network where results can be obtained analytically.

The remainder of this paper is organized as follows. In Sec.~\ref{sec:02} we present the equations of motion and illustrate the periodic and quasi-periodic dynamics that occur on undirected and directed networks. In Sec.~\ref{sec:03} we illustrate the occurrence of a Neimark-Sacker bifurcation and derive the genesis of quasi-periodic orbits. In Sec.~\ref{sec:04} we examine the onset of quasi-periodicity. In Sec.~\ref{sec:05} we consider the special case of a multi-neighbor ring where results can be obtained analytically. Finally, in Sec.~\ref{sec:06} we conclude with a discussion of our results.

\section{Equations of motion and network directedness}\label{sec:02}

A typical discrete-time random walk on a complex network describes the evolution of a probability vector $\bm{p}(t)$ given by
\begin{align}
p_i(t+1)=\sum_{j=1}^N\pi_{ij}p_j(t)\label{eq:01}
\end{align}
or in vector form
\begin{align}
\bm{p}(t+1)=\Pi\bm{p}(t)\label{eq:02}
\end{align}
where $\Pi$ is the transition matrix whose entry $\pi_{ij}$ describes the conditional probability of moving from node $j$ to node $i$ in one time step given that a random walker is currently at node $j$. To ensure that $\bm{p}(t)$ remains a probability, i.e., $\bm{p}(t)\in\Omega$ where
\begin{align}
\Omega = \{\bm{p}\in\mathbb{R}^N|p_i\ge0\text{ for }i=1,\dots,N,\text{ and }\sum_{i=1}^Np_i=1\},\label{eq:03}
\end{align}
we require $\Pi$ to be column stochastic so that the entries are non-negative and the columns sum to one. The entries $\pi_{ij}$ are themselves derived from the network structure using the adjacency matrix $A$ whose entry $a_{ij}$ describes the weight of the link from node $j$ to node $i$. In general, $A$ can be binary, i.e., entries take values $0$ if no link exists or $1$ if a link exists, or weighted, however here we will assume that no negatively weighted links exist. A network is undirected if for each link $j\to i$ an equal and opposite link $i\to j$ exists, so that $a_{ji}=a_{ij}$, or equivalently, $A^T=A$. An unbiased random walk on a network is described by transition probabilities $\pi_{ij}=a_{ij}/k_{j}^{\text{out}}$ where $k_{j}^{\text{out}}=\sum_{i=1}^Na_{ij}$ is the out-degree of node $j$. A random walk can be made biased by choosing transition probabilities to preferentially route random walkers towards or away from certain nodes, e.g., based on their degree.

Here we will study nonlinear random walks on complex networks by generalizing Eq.~(\ref{eq:02}) so that, rather than let $\Pi$ be static, it depends explicitly on the current system state, obtaining 
\begin{align}
\bm{p}(t+1) = \Pi(\bm{p}(t))\bm{p}(t).\label{eq:04}
\end{align}
In the most general framework a different function $f_{ij}$ can be used to define each entry of the transition matrix so that entries are given by
\begin{align}
\pi_{ij}(\bm{p})=\frac{a_{ij}f_{ij}(\bm{p})}{\sum_{l=1}^Na_{lj}f_{lj}(\bm{p})}.\label{eq:05}
\end{align}
Note that the mapping $\bm{F}(\bm{p})=\Pi(\bm{p})\bm{p}$ maps probability vectors to probability vectors, i.e., $F:\Omega\to\Omega$ provided that the functions $f_{ij}$ are positive, i.e., $f_{ij}:\Omega\to(0,\infty)$ and all nodes have positive degree. 

To model situations where random walkers are preferentially routed towards other nodes with relatively many or few other random walkers we consider the specific choice of an exponential biasing function with bias parameter $\alpha$, namely defining
\begin{align}
\pi_{ij}(\bm{p}) = \frac{a_{ij}\text{exp}(\alpha p_i)}{\sum_{l=1}^Na_{lj}\text{exp}(\alpha p_l)},\label{eq:06}
\end{align}
where negative (positive) values of $\alpha$ correspond to biasing random walkers towards nodes that themselves have few (many) random walkers. Together, Eqs.~(\ref{eq:04}) and (\ref{eq:06}) give the full system dynamics, which ultimately depends on the underlying network structure, encoded by the adjacency matrix $A$, and the bias parameter $\alpha$.

\begin{figure*}[t]
\centering
\epsfig{file =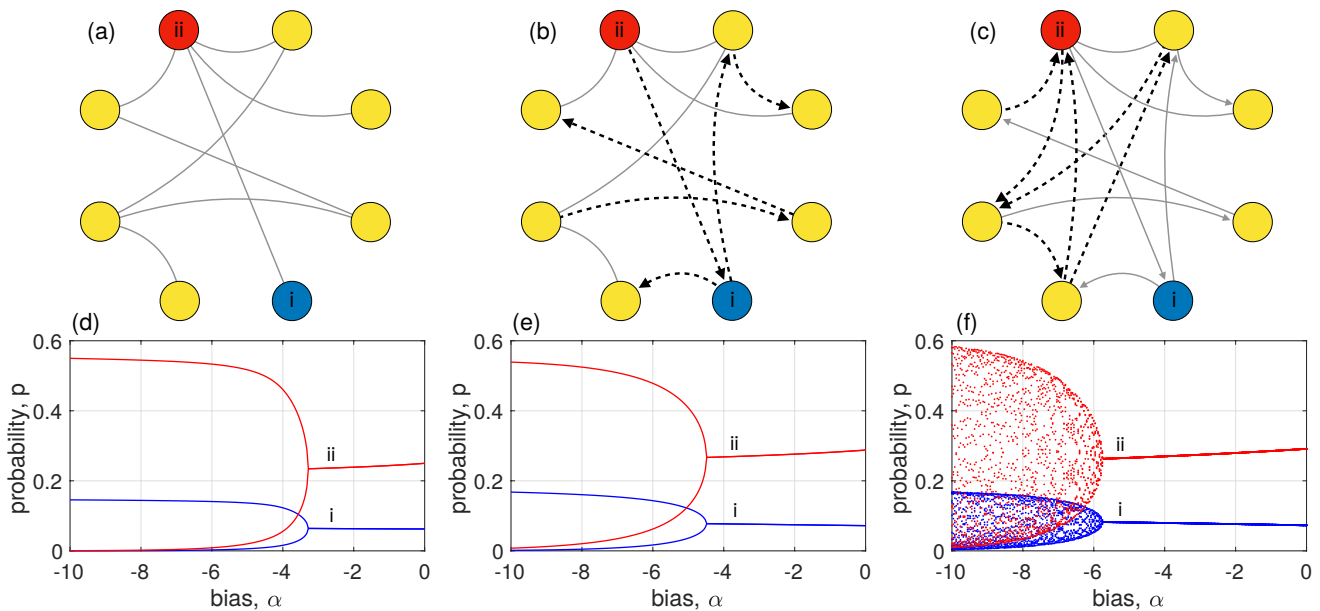, clip =,width=1.0\linewidth }
\caption{{\bf Nonlinear random walks and directedness}. (a)--(c) Three networks of size $N=8$ with mean degree $\langle k\rangle=2$: (a) is undirected, (b) is obtained by making three modifications (see text) to (a), and (c) is obtained by making three modifications to (c). In each case the new or modified links are highlighted as thick, dashed arrows. (d)--(f) As a function of the bias parameter $\alpha$, the long-term dynamics of the nonlinear random walk on networks illustrated in (a)--(c), respectively. For each value of $\alpha$ a transient of $10^4$ time steps is discarded and the probabilities of the nodes labelled ``i'' (blue) and ``ii'' (red) for the next 16 time steps are plotted.}\label{fig:01}
\end{figure*}

Our interest here lies in the nonlinear phenomenon that arises in the strongly nonlinear regime with negative bias and the dependence on network structure, specifically directedness. Recall that for typical undirected networks a period-doubling bifurcation occurs at a critical value of the bias parameter, $\alpha=\alpha_c$, beyond which the random walk converges to a period-two orbit. To explore the effect of directedness on nonlinear random walk dynamics we consider an illustrative example. In Fig.~\ref{fig:01}(a) we show an undirected network (size $N=8$ with mean degree $\langle k\rangle=N^{-1}\sum_{i=1}^Nk_i=2$). In the panel below, Fig.~\ref{fig:01}(d), we plot the steady-state probabilities corresponding to the blue and red nodes labelled ``i'' and ``ii'' for the nonlinear random walk as a function of $\alpha$. Note that a period-doubling bifurcation occurs at $\alpha_c\approx-3.28$ above and below which the long-term dynamics are period-one and period-two, respectively. 

Next, to explore the effect of directed network structures we consider structural modifications as follows. Interpreting each undirected link as a pair of equal and opposite directed links, we choose an undirected link at random, delete (at random) one of the directed counterparts, and create (at random) a new directed link elsewhere in the network between two previously disconnected nodes. We make three such modifications to the network illustrated in  Fig.~\ref{fig:01}(a), resulting in the network illustrated in Fig.~\ref{fig:01}(b). Each modified or new directed link is highlighted, plotted in a thick dashed curve with arrows indicating direction. Note that three modifications as described above results in six directed links in total -- three resulting from deleting half of an undirected link and three new directed links. In the panel below, Fig.~\ref{fig:01}(e), we again plot the steady-state probabilities corresponding to the blue and red nodes labelled ``i'' and ``ii'' for the nonlinear random walk as a function of $\alpha$. For the modified network the bifurcation is somewhat delayed, now occurring at $\alpha_c\approx-4.50$, as a period-doubling bifurcation that is qualitatively similar to the previous undirected network. 

Lastly, we make three more modification as described above to the network illustrated in Fig.~\ref{fig:01}(b), resulting in Fig.~\ref{eq:01}(c). Again we highlight each modified or new directed link is highlighted as a thick, dashed arrow. In the panel below, Fig.~\ref{fig:01}(f), we again plot the steady-state probabilities corresponding to the blue and red nodes labelled ``i'' and ``ii'' for the nonlinear random walk as a function of $\alpha$. For this new network structure, however, the results are qualitatively different than the previous networks. A bifurcation occurs at $\alpha_c\approx-5.75$ where the fixed point solution loses stability, however the dynamics beyond this point are not period-two, indicating that the bifurcation is no longer period-doubling. In particular, instead of two continuous branches signifying period-two orbits, the long-term dynamics populate the area within an envelope that is similar in shape to the period-two branches observed in the previous cases. 

\begin{figure}[t]
\centering
\epsfig{file =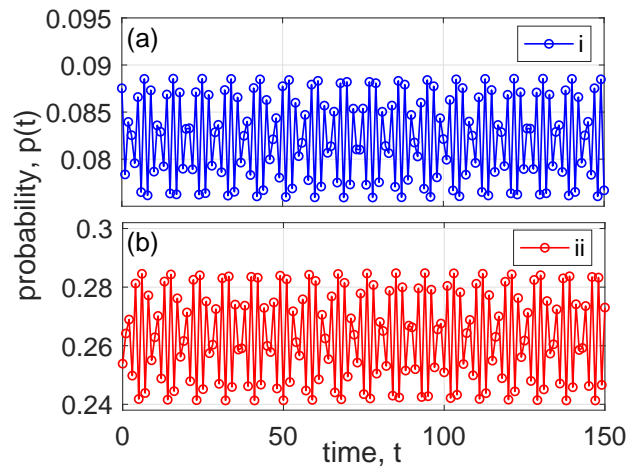, clip =,width=1.0\linewidth }
\caption{{\bf Quasi-periodic dynamics of nonlinear random walks}. Example time series of the probabilities labelled ``i'' (top) and ``ii'' (bottom) in the network illustrated in Fig.~\ref{fig:01}(c) for $\alpha = -5.76$, which is just beyond the bifurcation corresponding to the loss of stability of the fixed point.}\label{fig:02}
\end{figure}

Numerical simulations suggest that the long-term behavior of the nonlinear random walk on the network illustrated in Fig.~\ref{fig:01}(c) beyond the bifurcation is generically quasi-periodic. To see this, we plot in Fig.~\ref{fig:02} example time series of the long-term probabilities of nodes i and ii at $\alpha=-5.76$, just beyond the bifurcation. Note in particular that the time series appears to be {\it nearly} repeating itself after many iterations, but not exactly. In fact, analyzing longer time series than that shown here suggests that the dynamics are in fact not repeating themselves precisely, but returning to an arbitrarily close position over and over again, indicating quasi-periodic dynamics. (We also note here that straightforward numerical computations of Lyapunov exponents very near zero indicate that these dynamics are not chaotic.) As we will see below, this quasi-periodic behavior arises from a Neimark-Sacker bifurcation that gives rise to high-dimensional rotations in the space $\Omega$.

\section{Neimark-Sacker bifurcation and quasi-periodicity}\label{sec:03}

\begin{figure*}[t]
\centering
\epsfig{file =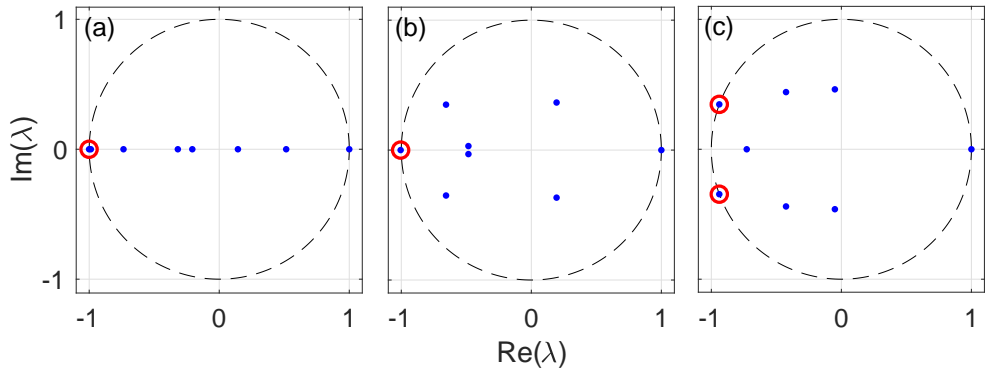, clip =,width=0.75\linewidth }
\caption{{\bf Spectra and bifurcations}. (a)--(c) For the networks illustrated in Fig.~\ref{fig:01}(a)--(c), respectively, the eigenvalue spectra of the Jacobian $DF(\bm{p}^*)$ in the complex plane at critical bias parameters $\alpha_c\approx-3.28$, $-4.50$, and $-5.75$. The unit circle is depicted as a dashed curve and the maximal nontrivial eigenvalues are highlighted using extra red circles.}\label{fig:03}
\end{figure*}

We now turn our attention to the bifurcations illustrated above. Regardless of the nature of the bifurcation for the specific case chosen, each such bifurcation corresponds to the loss of stability of the fixed point, i.e., stationary distribution, $\bm{p}^*$ that satisfies $\bm{p}^*=\Pi(\bm{p}^*)\bm{p}^*$. The stability of this fixed point is governed by the eigenvalues of the Jacobian $DF$ of the mapping $\bm{F}(\bm{p})=\Pi(\bm{p})\bm{p}$, given by
\begin{widetext}
\begin{align}
DF_{ij}(\bm{p}) = \left\{\begin{array}{rl}
\sum_{k=1}^N\frac{\alpha a_{ik}\text{exp}(\alpha p_i)\left\{\left[\sum_{l=1}^Na_{lk}\text{exp}(\alpha p_l)\right]-a_{ik}\text{exp}(\alpha p_i)\right\}}{\left[\sum_{l=1}^Na_{lk}\text{exp}(\alpha p_l)\right]^2}p_k&\text{ if }i=j,\\
\frac{a_{ij}\text{exp}(\alpha p_i)}{\sum_{l=1}^Na_{lj}\text{exp}(\alpha p_l)}-\alpha\sum_{k=1}^N\frac{a_{ik}a_{jk}\text{exp}(\alpha p_i)\text{exp}(\alpha p_j)}{\left[\sum_{l=1}^Na_{lj}\text{exp}(\alpha p_l)\right]^2}p_k&\text{ if }i\ne j.
\end{array}\right.
\end{align}
\end{widetext}
Specifically, the fixed point $\bm{p}^*$ is stable to perturbations in the space $\Omega$ if all eigenvalues $\lambda$ of $DF(\bm{p}^*)$ except for one are less than one in magnitude, i.e., satisfy $|\lambda|<1$. We note that one eigenvalue, which we denote $\lambda_1$, of $DF(\bm{p}^*)$ is always exactly one, i.e., $\lambda_1=1$, due to the conservation of probability of the mapping $\bm{F}(\bm{p})=\Pi(\bm{p})\bm{p}$. Thus, stability of $\bm{p}^*$ follows if all $|\lambda_j|<1$ for $j=2,\dots,N$ and the bifurcation corresponding to the loss of stability occurs when, as $\alpha$ is decreased, one or more eigenvalues meets and then exceeds $|\lambda_j|=1$.

To illustrate the nature of the bifurcations that we observe in the nonlinear random walk on different networks, we use the example networks illustrated in Fig.~\ref{fig:01}(a)--(c). In particular, we investigate the spectrum of the Jacobian $DF(\bm{p}^*)$ at the bifurcation points, observed at $\alpha_c\approx-3.28$, $-4.50$, and $-5.75$, respectively. In Fig.~\ref{fig:03}(a)--(c) we plot the spectra of eigenvalues of the three cases, respectively, in the complex plane. Note that each spectrum contains exactly one eigenvalue $\lambda_1=1$, as discussed above. Ignoring this trivial eigenvalue, we highlight the critical eigenvalue(s) of maximal magnitude for each case with an additional red circle. Importantly, we see that for both cases (a) and (b) there is a single maximal eigenvalue at precisely $\lambda=-1$. This corresponds to a classical period-doubling bifurcation, as the fixed point losses stability, specifically giving rise to period-two oscillations that alternate above and below the fixed point and agreeing with the dynamics we see in Figs.~\ref{fig:01}(d) and (e). Case (c), however, presents a critical difference in that, rather than there being a critical eigenvalue at $\lambda=-1$, there are two complex conjugate maximal eigenvalues at $\lambda=\text{exp}(\pm i\theta)$ (with $0<\theta<\pi$), which indicates a Neimark-Sacker bifurcation~\cite{Kuznetsov2010}. These complex-values critical eigenvalues naturally give rise to behavior just beyond the bifurcation that does not simply oscillate in a period-two orbit, but rather rotate in a high-dimensional space. These rotations are precisely what is observed in Fig.~\ref{fig:01}(f), although this can be seen better from an analysis we will present below. We note that while the non-critical eigenvalues of $DF(\bm{p}^*)$ for case (b) are also complex, it is the maximal eigenvalues that give rise to dynamics just beyond the bifurcation, explaining why the dynamics from case (b) are in fact period-two.

To shed more light on the nature of the rotations that arise in the case of a Neimark-Sacker bifurcation, we consider perturbations to the fixed point of the form $\bm{p}(t)=\bm{p}^*+\delta\bm{p}(t)$, where $\|\delta\bm{p}(t)\|\ll1$ (where, since we are considering probability vectors, we use the $\ell^1$ norm). Inserting this into the map, we have that
\begin{align}
\bm{p}^*+\delta\bm{p}(t+1)&=\bm{F}(\bm{p}^*+\delta\bm{p}(t))\label{eq:08}\\
&=\bm{F}(\bm{p}^*)+DF(\bm{p}^*)\delta\bm{p}(t)+\mathcal{O}(\|\delta \bm{p}(t)\|^2).\label{eq:09}
\end{align}
Since $\bm{F}(\bm{p}^*)=\bm{p}^*$ this reduces to 
\begin{align}
\delta\bm{p}(t+1)=DF(\bm{p}^*)\delta\bm{p}(t)+\mathcal{O}(\|\delta \bm{p}(t)\|^2),\label{eq:10}
\end{align}
from which we can see explicitly that the fixed point becomes unstable via a growing perturbation if one or more eigenvalues of $DF(\bm{p}^*)$ is larger than one in magnitude. In particular, $DF(\bm{p}^*)$ has a number of important spectral properties which we will take advantage of. Like $\Pi(\bm{p}^*)$, the columns of $DF(\bm{p}^*)$ sum to one, which implies two important things. First, there is always an eigenvalue $\lambda_1=1$ (whose corresponding left eigenvector is constant) and whose corresponding right eigenvector $\bm{v}^1$ generically has a non-zero sum (and typically has all positive entries). Second, and more importantly, all other eigenvectors, i.e., eigenvectors corresponding to eigenvalues $\lambda_i\ne1$ for $i=2,\dot,N$, sum to zero. In particular, this implies that any perturbation $\delta\bm{p}$ of which we are interested that conserves the sum $\sum_{i=1}^N(p_i+\delta p_i)=1$ has no component in the direction of $\bm{v}^1$, and therefore can be written only as a linear combination of the other eigenvectors, namely 
\begin{align}
\delta\bm{p}(0)=\sum_{j=2}^N\beta_j\bm{v}^j.\label{eq:11}
\end{align}
(Note that to ensure that $\|\delta\bm{p}(0)\|\ll1$ we need $\beta_j\ll1$.) Inserting this into Eq.~(\ref{eq:10}) yields, after $t$ time steps,
\begin{align}
\delta\bm{p}(t)=\sum_{j=2}^N\beta_j\lambda_j^t\bm{v}^j.\label{eq:12}
\end{align}
In particular, when $\alpha>\alpha_c$ and the stationary distribution is linearly stable the relevant eigenvalues are strictly less than one, i.e., $|\lambda_j|<1$ for $j=2,\dots,N$ so that all eigenmodes in Eq.~(\ref{eq:12}) decay to zero as $t\to\infty$. However, when $\alpha=\alpha_c$ and the first pair of complex-conjugate eigenvalues are precisely one in magnitude, the modes in the corresponding eigenvector directions do not decay. Denoting these two eigenvalues $\lambda_2$ and $\lambda_3$, where $\lambda_3=\overline{\lambda}_2$ and $\bm{v}_3=\overline{\bm{v}}_2$, we have that, for sufficiently large $t$, 
\begin{align}
\delta\bm{p}(t)\to\beta_2\lambda_2^t\bm{v}_2+\beta_3\lambda_3^t\bm{v}_3.\label{eq:13}
\end{align}
Note also that, since the perturbation $\delta\bm{p}(0)$ is real, we have that $\beta_3=\overline{\beta}_2$. Since $|\lambda_{2,3}|=1$, we may write $\lambda_{2,3}=e^{\pm i\theta}$, which, after inserting this into Eq.~(\ref{eq:13}), yields rotations of infinitesimal size given by
\begin{align}
\delta\bm{p}(t)=2(\mu\bm{u}-\nu\bm{w})\cos(t\theta)-2(\mu\bm{w}+\nu\bm{u})\sin(t\theta),\label{eq:14}
\end{align}
where $\bm{v}_{2,3}=\bm{u}\pm i\bm{w}$ and $\beta_{2,3}=\mu\pm i\nu$. In particular, Eq.~(\ref{eq:14}) reveals high-dimensional rotations in $\Omega$ that take place around the fixed point $\bm{p}^*$. In general, provided that the argument $\theta/\pi$ is irrational, then since $t$ takes only integer values the rotations described in Eq.~(\ref{eq:14}) yield quasi-periodic dynamics, never repeating themselves exactly, but coming arbitrarily close infinitely many times. Moreover, it can be observed numerically that decreasing $\alpha$ below the critical bifurcation value $\alpha_c$ yields similar rotational dynamics as predicted at the bifurcation $\alpha=\alpha_c$ but with amplitudes that grow from an initially small perturbation and saturate due to nonlinear effects. This saturation is observed as the envelope in Fig.~\ref{fig:01}(f). These dynamics can be thought of as a generalization to the alternations of period-two dynamics in that, if $\theta$ is close to $\pi$ the dynamics alternate, but with with a slow rotation of angle $\pi-\theta$ each iteration.

\section{Onset of Quasi-periodicity}\label{sec:04}

The analysis presented above demonstrates how quasi-periodicdynamics emerge when the stationary state loses stability due to a Neimark-Sacker bifurcation. However, the discrete network perturbations (i.e., link rewirings) used in the previous section do little to illuminate the onset of quasi-periodicity as a transition away from periodicity. To better understand this process we consider a network model where the directedness of a network can be varied continuously. 

In this vein, we begin with an undirected, binary network of $N$ nodes with adjacency matrix $A$, and we construct two additional $N\times N$ matrices $B$ and $C$. Assuming $A$ has $M$ undirected, unweighted links, and thus $2M$ non-zero (one) entries, we populate both $B$ and $C$ with $M$ entries of $1$. Critically, each non-zero entry of $B$ and $C$ is opposite to a zero entry, so if $b_{ij}=1$ then $b_{ji}=0$ and if $c_{ij}=1$, then $c_{ji}=0$. Moreover, the non-zero entries of $B$ correspond to entries where no link exist in $A$, i.e., $a_{ij}=a_{ji}=0$ and the non-zero entries of $C$ correspond to entries where a link does exist in $A$, i.e., $a_{ij}=a_{ji}=1$. Thus, by adding $B$ to $A$ we introduce new directed links in spots where no link existed previously, and by subtracting $C$ from $A$ we remove one directed half of an undirected link initially in $A$. Note that since $B$ and $C$ contain $M$ non-zero entries each, $M$ new directed links are created by $B$ and one half of each of the $M$ originally undirected links in $A$ are removed by subtracting $C$. Finally, in order to obtain a network with truly continuously-varying directedness, we consider the new adjacency matrix $A(\epsilon)$ given by
\begin{align}
A(\epsilon)=A+\epsilon(B-C),\label{eq:15}
\end{align}
where $\epsilon\in[0,1]$ is the directedness parameter with $\epsilon=0$ and $1$ corresponding to, respectively, the original undirected network and a new network whose directedness is maximal in the sense that no non-zero entry of $A(\epsilon)$ has an opposite non-zero counterpart and the mean-degree of $A(\epsilon)$ is conserved.

\begin{figure}[t]
\centering
\epsfig{file =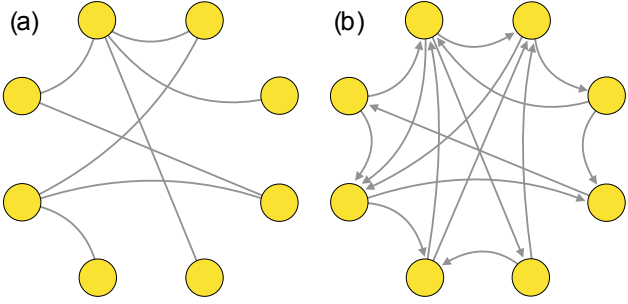, clip =,width=1.0\linewidth }
\caption{{\bf Limiting networks}. The limiting network structures represented by adjacency matrices $A(\epsilon)$ for $\epsilon=0$ and $1$ [(a) and (b), respectively].}\label{fig:04}
\end{figure}

\begin{figure}[t]
\centering
\epsfig{file =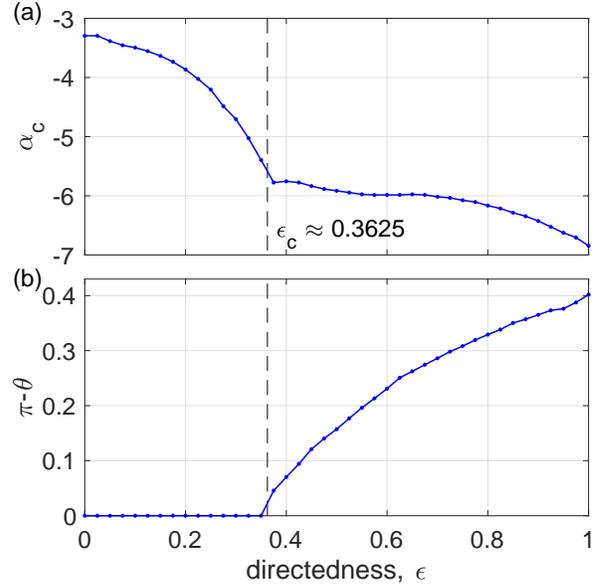, clip =,width=1.0\linewidth }
\caption{{\bf Onset of quasi-periodicity}. For the example network $A(\epsilon)$ for $\epsilon=0$ and $1$ illustrated in Fig.~\ref{fig:04}, (a) the critical value $\alpha_c$ and (b) the angle off the negative real branch $\pi-\theta$ of the critical eigenvalue(s) at the bifurcation corresponding to the loss of stability of the stationary distribution. The onset of quasi-periodicity is denoted with a vertical dashed line at $\epsilon\approx0.3625$.}\label{fig:05}
\end{figure}

To explore the onset of quasi-periodicity we consider the undirected network illustrated in Fig.~\ref{fig:04}(a), giving the adjacency matrix $A$, along with rewiring matrices $B$ and $C$ that yield, at $\epsilon=1$, the directed network illustratted in Fig.~\ref{fig:04}(b). (Note that the undirected network is identical to that illustrated in Fig.~\ref{fig:01}(a).) We then vary $\epsilon$ form zero to one, for each value slowly decreasing $\alpha$ as the dynamics simulated until we reach the bifurcation at $\alpha=\alpha_c$ defined by the first eigenvalue or pair of eigenvalues of $DF(\bm{p}^*)$ crossing the complex unit circle. Once at the bifurcation, we then calculate the phase angle(s) $\theta$ of the critical eigenvalue(s), taking the positive angle in the range $(-\pi,\pi]$ when eigenvalues come in complex conjugate pairs. In Fig.~\ref{fig:05} we show the results of this numerical exploration, plotting in panels (a) and (b) the critical bias parameter $\alpha_c$ and the angle of the phase off the negative real axis $\pi-\theta$, respectively, as a function of the directedness parameter $\epsilon$. In particular, we observe that when $\epsilon$ surpasses $\epsilon_c\approx0.3625$ the critical eigenvalues passing through the complex unit circle begin to come in complex-conjugate pairs with $\theta\in(0,\pi)$. As $\epsilon$ is continuously moved through this critical value, two real, negative eigenvalues (i.e., with $\theta=\pi$) collide at $\lambda=-1$ and move off the $\theta=\pi$ branch into the positive and negative imaginary halves of the complex plane. For this particular example $\theta$ continues to move off away from the $\theta=\pi$ branch as $\epsilon$ is further increased, which appears to be typical in our other numerical explorations (not shown). Finally, since this phenomenon occurs at $\alpha=\alpha_c$, the onset of quasi-periodicity represents a codimenson-two point at $(\alpha,\epsilon)=(\alpha_c,\epsilon_c)$.

\section{Special case: the four-neighbor ring}\label{sec:05}

We conclude by considering a case where the dynamics can be described analytically. In particular, we consider the case of a four-neighbor ring, i.e., a ring where each of the $N$ nodes is connected to each of its two nearest neighbors on each side. (We chose the four-neighbor ring so that the network is primitive, which is not the case for the typical two-neighbor ring.) To generate a directed network structure we introduce a parameter $\epsilon$ that weights the links with a chosen orientation. In particular, indexing the nodes in order around the ring, we let $a_{ij}=1+\epsilon$ if $i=j+1$ or $j+2$, $a_{ij}=1-\epsilon$ if $i=j-1$ or $j-2$, and otherwise $a_{ij}=0$.

We now perform a linear stability analysis by seeking the eigenvalues of the Jacobian $DF(\bm{p}^*)$, whose eigenvalue equation $DF(\bm{p}^*)\bm{v}=\lambda\bm{v}$ implies, for all $i=1,\dots,N$,
\begin{align}
\lambda v_i=\sum_{j=i-4}^{i+4}DF_{ij}v_{j},\label{eq:16}
\end{align}
where periodic indexing $N+k\mapsto k$ and $-k\mapsto N+1-k$ is assumed. Moreover, since nodal in- and out-degrees are identical for throughout the network we have that the stationary distribution is given by the constant vector $\bm{p}^*=\bm{1}/N$, yielding entries of $DF(\bm{p})$ given by
\begin{align}
DF_{ij}(\bm{p}^*)=\left\{\begin{array}{rl}
-\frac{\alpha(1-\epsilon^2)}{16N}&\text{if }j=i-4,\\
-\frac{2\alpha(1-\epsilon^2)}{16N}&\text{if }j=i-3,\\
\frac{(1-\epsilon)}{4}-\frac{\alpha(1-\epsilon^2)}{16N}&\text{if }j=i-2,\\
\frac{(1-\epsilon)}{4}-\frac{2\alpha(1+\epsilon^2)}{16N}&\text{if }j=i-1,\\
\frac{\alpha}{N}\left(1-\frac{1+\epsilon^2}{4}\right)&\text{if }j=i,\\
\frac{(1+\epsilon)}{4}-\frac{2\alpha(1+\epsilon^2)}{16N}&\text{if }j=i+1,\\
\frac{(1+\epsilon)}{4}-\frac{\alpha(1-\epsilon^2)}{16N}&\text{if }j=i+2,\\
-\frac{2\alpha(1-\epsilon^2)}{16N}&\text{if }j=i+3,\\
-\frac{\alpha(1-\epsilon^2)}{16N}&\text{if }j=i+4.
\end{array}\right.\label{eq:17}
\end{align}
The rotational symmetry and periodic nature of the eigenvector equations suggest that eigenvectors come in the form of $v_j=\text{exp}(ij\phi)$ for $\phi=2\pi k/N$ for wavenumbers $k=0,\dots,N-1$. Inserting this into Eq.~(\ref{eq:16}) yields the following explicit formula for the eigenvalue $\lambda$:
\begin{widetext}
\begin{align}
\lambda&=\sum_{j=-4}^4DF_{i,i+j}\text{exp}(ij\phi)=\frac{\alpha}{N}\left(1-\frac{1+\epsilon^2}{4}\right)\nonumber\\
&-\frac{\alpha(1-\epsilon^2)}{16N}e^{-4i\phi}-\frac{2\alpha(1-\epsilon^2)}{16N}e^{-3i\phi}+\left[\frac{(1-\epsilon)}{4}-\frac{\alpha(1-\epsilon^2)}{16N}\right]e^{-2i\phi}+\left[\frac{(1-\epsilon)}{4}-\frac{2\alpha(1+\epsilon^2)}{16N}\right]e^{-i\phi}\nonumber\\ 
& + \left[\frac{(1+\epsilon)}{4}-\frac{2\alpha(1+\epsilon^2)}{16N}\right]e^{i\phi} + \left[\frac{(1+\epsilon)}{4}-\frac{\alpha(1-\epsilon^2)}{16N}\right]e^{2i\phi} -\frac{2\alpha(1-\epsilon^2)}{16N}e^{3i\phi}-\frac{\alpha(1-\epsilon^2)}{16N}e^{4i\phi}.\label{eq:18}
\end{align}
\end{widetext}
Thus, Eq.~(\ref{eq:18}) gives an eigenvalue $\lambda$ for each $\phi=2\pi k/N$ for $k=0,\dots,N-1$, thereby providing the full eigenvalue spectrum for $DF(\bm{p}^*)$. Thus, a linear stability analysis can be done for any four-neighbor ring of size $N$ by decreasing $\alpha$ until the first single or pair of eigenvalues given by Eq.~(\ref{eq:18}) surpasses $|\lambda|=1$ and then inspecting the properties of the resulting critical eigenvalue(s).

\begin{figure}[t]
\centering
\epsfig{file =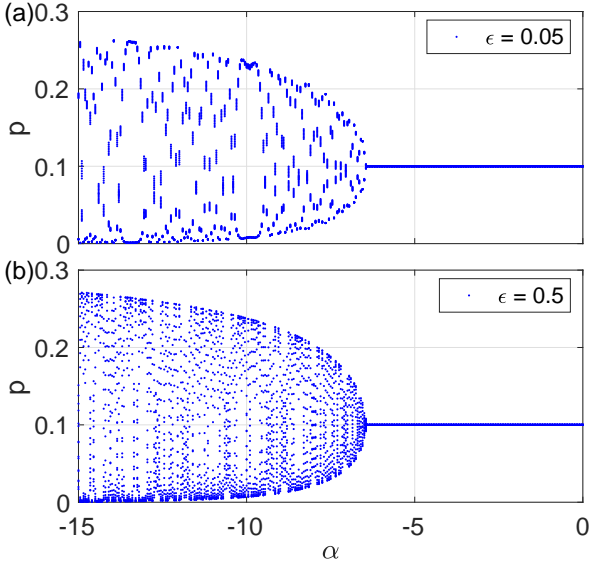, clip =,width=1.0\linewidth }
\caption{{\bf Nonlinear random walk on the four-neighbor ring}. As a function of the bias parameter $\alpha$, the long-term dynamics of the nonlinear random walk on the four-neighbor ring of size $N=10$ with directedness parameters $\epsilon=0.05$ (a) and $0.5$ (b). For each value of $\alpha$ a transient of $10^4$ time steps is discarded and the probabilities of a randomly chosen node for the next 16 time steps are plotted.}\label{fig:06}
\end{figure}

\begin{figure}[t]
\centering
\epsfig{file =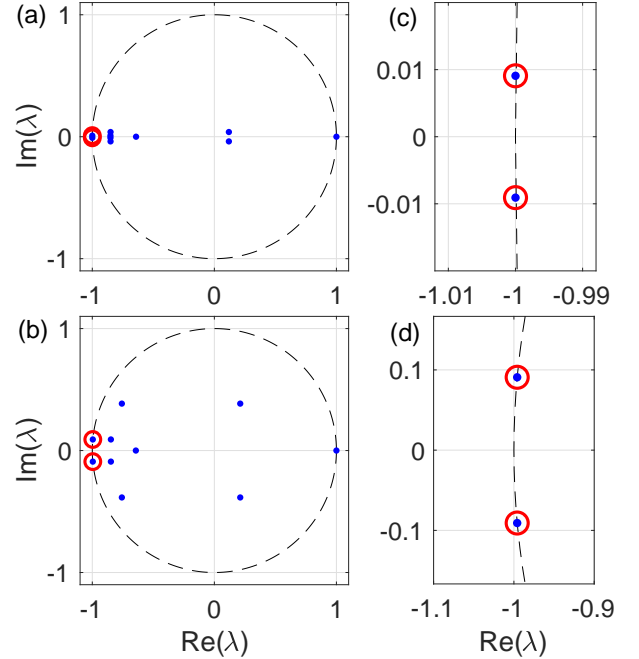, clip =,width=1.0\linewidth }
\caption{{\bf Eigenvalue spectrum for the four-neighbor ring}. (a) and (b): For the four-neighbor ring of size $N=10$ with directedness $\epsilon=0.05$ and $0.5$, respectively, the eigenvalue spectrum of the Jacobian $DF(\bm{p}^*)$ at critical values $\alpha_c\approx-6.415$ and $-6.432$. The unit circle is depicted as a dashed curve and the maximal nontrivial eigenvalues are highlighted using extra red circles. (c) and (b): Zoomed-in view on the maximal nontrivial eigenvalues.}\label{fig:07}
\end{figure}

In Fig.~\ref{fig:06} we plot the results from numerical simulations of a four-neighbor ring of size $N=10$ with directedness parameters $\epsilon=0.05$ and $0.5$ in panels (a) and (b), respectively. at each value of $\alpha$ we simulate through a transient of $10^4$ iterations and plot the resulting probabilities for a randomly chosen node (since all nodes are topologically equivalent) for the next $16$ iterates. The results appears very similar for the two cases, except for the ``streakyiness'' present for $\epsilon=0.05$. This can be explained by inspecting spectrum of the Jacobian at the bifurcation, which for the two cases occurs at $\alpha_c\approx-6.415$ and $-6.432$, respectively, according to our predictions using Eq.~(\ref{eq:18}). In Fig.~\ref{fig:07} we plot the spectrum $DF(\bm{p}^*)$ for each case in panels (a) and (b), respectively. The critical eigenvalues, circled in red, are in fact all complex, however for $\epsilon=0.05$ they lie just off of the negative real axis ($\pi-\theta\approx 0.00919$), indicating that the dynamics are comprised of a slow rotation about an alternation. For $\epsilon=0.5$ the angle off of the negative real axis is significantly greater ($\pi-\theta\approx0.0919$), yielding rotations that are not nearly as slow, thereby filling a larger amount of space in the 16 iterations plotted in Fig.~\ref{fig:6}. In panels (c) and (d) we zoom-in on the critical eigenvalues for a better view (and to ensure that the eigenvalues for $\epsilon=0.05$ are in fact complex).

\begin{figure}[t]
\centering
\epsfig{file =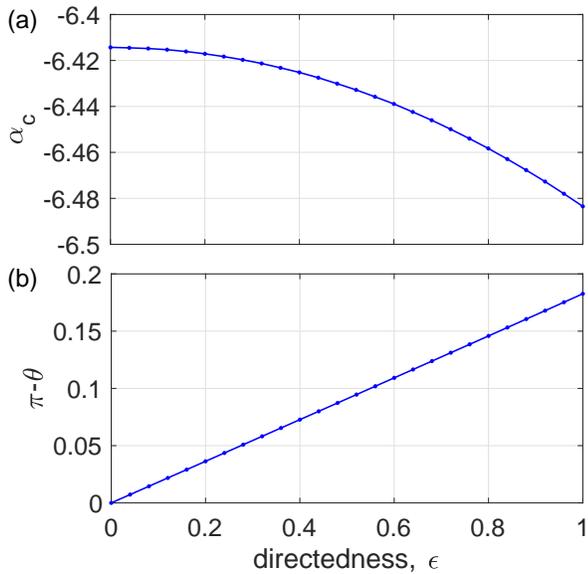, clip =,width=1.0\linewidth }
\caption{{\bf Onset of quasi-periodicity for the four-neighbor ring}. For the four-neighbor ring as a function of directedness $\epsilon$, (a) the critical value $\alpha_c$ and (b) the angle off the negative real branch $\pi-\theta$ of the critical eigenvalue(s) at the bifurcation corresponding to the loss of stability of the stationary distribution. The onset of quasi-periodicity occurs at $\epsilon=0$.}\label{fig:08}
\end{figure}

Lastly, we inspect the overall system dynamics as a function of $\epsilon$ by plotting in Figs.~\ref{fig:08}(a) and (b), respectively, the critical bifurcation value $\alpha_c$ and the angle of the critical eigenvalue off the negative real axis $\pi-\theta$ as the directedness $\epsilon$ is increased. Most notably, the angle offset $\pi-\theta$ differs from zero for any non-zero directedness value, indicating that the onset of quasi-periodicity actually does occur at $\epsilon=0$. This is unlike the example network used above (see Fig.~\ref{fig:04}) where the onset occurred at a finite value of $\epsilon$.

\section{Discussion}\label{sec:06}

In this paper we have investigated the dynamics that arise in nonlinear random walks on complex networks as network topologies become more and more directed. In particular, we have shown that the loss of stability of the stationary distribution may give rise to quasi-periodic dynamics if the network structure is sufficiently directed. This is in contrast to undirected networks and some ``weakly'' directed networks where this bifurcation gives rise to period-two orbits via a period-doubling bifurcation. In particular, the quasi-periodic dynamics observed for directed network arises from a Neimark-Sacker bifurcation where, rather than a single eigenvalue passing through $-1$, a pair of complex conjugate eigenvalues pass through the unit circle, giving rise to high-dimensional rotations about the unstable stationary distribution.

We have also investigated the onset of quasi-periodicity in terms of tuning the directedness of a network. In particular, at the onset we see that the Neimark-Sacker bifurcation occurs as two real-valued eigenvalues have angle close to $\pi$, which manifest in alternations with a slow rotation. Finally we showed that in some cases analytical results are attainable, for instance the four-neighbor ring analyzed above.



\bibliographystyle{plain}

\begin{thebibliography}{99}
\bibitem{Noh2002PRL} J. D. Noh and H., Rieger, Random walks on complex networks, Phys. Rev. Lett. {\bf 92}, 118701 (2004).
\bibitem{Masuda2017PR} N. Masuda, M. A. Porter, and R. Lambiotte, Random walks and diffusion on networks, Phys. Rep. {\bf 716-717}, 1 (2017).

\bibitem{Brin1998} S. Brin and L. Page, The anatomy of a large-scale hypertextual web search engine. In: Computer Networks and ISDN Systems. Proc. of the Seventh International World Wide Web Conference, {\bf 30}, 107 (1998).
\bibitem{Page1999} L. Page, S. Brin, R. Motwani, and T. Winograd, The pagerank citation ranking: bringing order to the web. Technical Report 1999-66, Stanford InfoLab (1999).
\bibitem{Gleich2015SIAM} D. F. Gleich, Pagerank beyond the web, SIAM Rev. {\bf 57}, 321 (2015).

\bibitem{GomezGardenes2008PRE} J. G\'{o}mez-Garde\~{n}es and V. Latora,  Entropy rate of diffusion processes on complex networks, Phys. Rev. E {\bf 78}, 065102(R) (2008).
\bibitem{Sinatra2011PRE} R. Sinatra, J. G\'{o}mez-Garde\~{n}s, R. Lambiotte, V. Nicosia, and V. Latora, Maximal-entropy random walks in complex networks with limited information, Phys. Rev. E {\bf 83}, 030103(R) (2011).

\bibitem{Gorenflo2002} R. Gorenflo, F. Mainardi, D. Moretti, G. Pagnini, and P. Paradisi, Discrete random walk models for space–time fractional diffusion, Chem. Phys. {\bf 284}, 521 (2002).
\bibitem{Nicosia2017PRL} V. Nicosia, P. S. Skardal, A. Arenas, and V. Latora, Collective phenomena emerging from the interactions between dynamical processes in multiplex networks, Phys. Rev. Lett. {\bf 118}, 138302 (2017).

\bibitem{Rosvall2008PNAS} M. Rosvall, and C. T.  Bergstrom, Maps of random walks on complex networks reveal community structure, Proc. Natl. Acad. Sci. {\bf 105}, (2008).
\bibitem{Asllani2018PRL} M. Asllani, T. Carletti, F. Di Patti, D. Fanelli, and F. Piazza, Hopping in the crowd to unveil network topology, Phys. Rev. Lett. {\bf 120}, 158301 (2018).

\bibitem{Coifman2005PNAS} R. R. Coifman, S. Lafon, A. B. Lee, M. Maggioni, B. Nadler, F. Warner, and S. W. Zucker, Geometric diffusion as a tool for harmonic analysis and structure definition of data: Diffusion maps, Phys. Rev. Lett. {\bf 102}, 7426 (2005).

 \bibitem{Durrett} R. Durrett and R. Durrett, {\it Essentials of Stochastic Processes} (Springer, Berlin, 2016).
 \bibitem{MacCluer2000SIAM} C. R. MacCluer, The many proofs and applications of Perron’s theorem, SIAM Rev. {\bf 42}, 487 (2000).
 
 \bibitem{Skardal2019JNS} P. S. Skardal and S. Adhikari, Dynamics of nonlinear random walks on complex networks, J. Nonlinear Sci. {\bf 29}, 1419 (2019).
 
 \bibitem{Kolokoltsov2010} V. N. Kolokoltsov, {\it Nonlinear Markov Processes and Kinetic Equations} (Cambridge University Press, Cambridge, 2010).
\bibitem{Frank2013} T. Frank, Strongly nonlinear stochastic processes in physics and the life sciences, ISRN Math. Phys. {\bf 2013},149169 (2013).

\bibitem{Sacker2009} R. J. Sacker, Introduction to the 2009 re-publication of the `Neimark-Sacker bifurcation theorem', J. Differ. Equ. Appl. {\bf 15}, 753 (2009).
\bibitem{Kuznetsov2010} Y. A. Kuznetsov, {\it Elements of Applied Bifurcation Theory} (Springer, New York, 2010)


%
%

\end{thebibliography}

\end{document}